\documentclass[12pt]{article}
\usepackage{amsmath,amssymb,graphicx,slashbox}
\usepackage{color}

\def\mydate{27 November 2010}
\def\ignore#1{{}}

\ignore{
\textheight=24.5cm
\textwidth=16cm
\topmargin=-1.5cm
\oddsidemargin=0.0cm
\evensidemargin=0.0cm
}

\newcommand{\bea}{\begin{eqnarray}}
\newcommand{\eea}{\end{eqnarray}}

\makeatletter
\@addtoreset{equation}{section}
\makeatother

\newcommand{\beeq}{\begin{equation}}
\newcommand{\eneq}{\end{equation}}
\newcommand{\beqn}{\begin{eqnarray}}
\newcommand{\eeqn}{\end{eqnarray}}

\def\mybig{\displaystyle \strut }

\def\dd{\partial}
\def\la{\raise.16ex\hbox{$\langle$}\lower.16ex\hbox{}  }
\def\ra{\raise.16ex\hbox{$\rangle$}\lower.16ex\hbox{} }
\def\go{\rightarrow}

\def\onehalf{ \hbox{${1\over 2}$} }

\def\Tr{{\rm Tr \,}}
\def\tr{{\rm tr \,}}
\def\eff{{\rm eff}}

\def\EM{{\rm EM}}
\def\diag{{\rm diag ~}}

\def\KK{{\rm KK}}
\def\bulk{{\rm bulk}}
\def\brane{{\rm brane}}

\def\ep{\epsilon}
\def\psibar{ \psi \kern-.65em\raise.6em\hbox{$-$} }
\def\psibarl{ \psi \kern-.65em\raise.6em\hbox{$-$} \lower.6em\hbox{} }

\def\myeq{\!\!\!=\!\!\!}
\def\myfrac#1#2{{\mybig #1\over \mybig #2}}

\begin{document}

\thispagestyle{empty}

{\small \noindent \mydate    \hfill OU-HET 681/2010}

{\small \noindent ver. 2  }

\vspace{4.0cm}

\baselineskip=35pt plus 1pt minus 1pt

\begin{center}
{\LARGE \bf 
$H$ parity and the stable Higgs boson\\
in the $SO(5) \times U(1)$ gauge-Higgs unification 
}
\end{center}

\vspace{2.5cm}
\baselineskip=20pt plus 1pt minus 1pt

\begin{center}
{\bf  
Yutaka Hosotani, Minoru Tanaka and Nobuhiro Uekusa
}


{\small \it Department of Physics, 
Osaka University, 
Toyonaka, Osaka 560-0043
Japan} \\
\end{center}


\vskip 3.0cm
\baselineskip=20pt plus 1pt minus 1pt

\begin{abstract}
In  the $SO(5) \times U(1)$ gauge-Higgs unification model 
in the Randall-Sundrum warped space there results the conservation of  the $H$ parity.  
The $H$ parity is assigned to all  4D fields including  excited modes in Kaluza-Klein 
towers.  
The neutral Higgs boson is the lightest particle of odd $H$ parity, 
consequently becoming absolutely stable.  Its  mass is found to be
$70 \sim135 \,$GeV for the warp factor  $z_L = 10^5 \sim 10^{15}$.
\end{abstract}



\newpage

\baselineskip=20pt plus 1pt minus 1pt

\section{Introduction}
The Higgs boson is the only particle yet  to be found 
in the standard model of electroweak interactions.  
It is not clear, however,  if the Higgs boson appears as described in the 
standard model (SM).
New physics may be hidden behind it.

One possible scenario is gauge-Higgs unification, in which spacetime has 
more than four dimensions and electroweak  gauge symmetry is  broken
by quantum dynamics in the extra dimension.\cite{YH1, Davies1, YH2}
The 4D Higgs boson, which becomes a part of gauge fields, 
appears as an Aharonov-Bohm (AB) phase in a 
non-simply-connected extra dimension.  
Its finite mass $m_H$  is generated at the quantum level.
A non-vanishing AB phase $\theta_H$,
or the Higgs vev, induces electroweak symmetry breaking and gives masses
to quarks, leptons, $W$ and $Z$.\cite{Hatanaka1998}-\cite{Alves2010}

In the $SO(5) \times U(1)$ gauge-Higgs unification model it has been shown
that the value $\theta_H = \onehalf \pi$ is dynamically chosen,\cite{HOOS} 
and the 4D Higgs boson  becomes stable.\cite{HKT}  It has been shown that
a new parity, $H$ parity, appears among low energy particles.
Only the Higgs boson is $H$ parity odd, while all other particles in the 
standard model are $H$ parity even.  The stability implies that Higgs bosons
become dark matter of the universe.  The relic density of 
cold dark matter observed at WMAP can be obtained with $m_H \sim 70\,$GeV.

The gauge-Higgs unification scenario leads to many phenomenological
consequences.  The nature of the Higgs boson as an AB phase leads to
the stability against quantum corrections which gives a solution to the 
gauge-hierarchy problem.\cite{Hatanaka1998}
Gauge-couplings of quarks and leptons slightly deviate from those
in SM, whereas  significant deviation appears in the Higgs couplings.\cite{HS2, HK, HNU}
Distinctive prediction for anomalous magnetic moment and electric dipole moment 
has been  discussed.\cite{Lim2007a, Lim2007b, Lim2009}   
The spectrum and couplings of Kaluza-Klein (KK) excited
states may differ from those in other extra-dimensional theories such 
as UED models.

In this paper we focus on the Higgs boson in the gauge-Higgs unification.  
As mentioned above,  the Higgs boson becomes stable in a class of 
the $SO(5) \times U(1)$ gauge-Higgs unification models as a result of the 
$H$ parity conservation.
In this regards we note that stable, 
or almost stable,  Higgs bosons have appeared in other models.
The inert doublet Higgs model of Deshpande and Ma is among them, 
in which a second  Higgs field is  introduced in addition to 
the standard Higgs field giving masses to quarks, leptons, $W$ and 
$Z$.\cite{Ma1978}  
The model has a $Z_2$ symmetry such that the second Higgs field is odd, 
while other low energy fields are even.
Because of this new $Z_2$ symmetry, or parity, the lightest Higgs boson of
odd parity becomes stable.  Many implications to dark matter and
neutrino physics have been discussed.\cite{Ma2006}-\cite{Yaguna2010}
Similarly the inert triplet Higgs model also serves as a minimal dark 
matter model.\cite{Cirelli2006}-\cite{Perez2009}

Although there is similarity in the Higgs boson between the inert Higgs models and 
the gauge-Higgs unification, there is crucial difference.
In the $SO(5) \times U(1)$ gauge-Higgs unification there is only one
Higgs doublet which is responsible for symmetry breaking and mass generation,
and at the same time becomes absolutely stable.
The $Z_2$ parity in the inert Higgs model is introduced by hand,
whereas the $H$ parity in the gauge-Higgs unification is hidden
in the original minimal model.  It dynamically emerges as a result of
the fact that the AB phase $\theta_H=\onehalf \pi$  is realized
in the vacuum.

Dynamically emergent $H$ parity plays a key role for the stability of the 
Higgs boson.  Previously $H$ parity has been assigned only for low 
energy fields in the $SO(5) \times U(1)$ gauge-Higgs unification model. 
In this paper we show that the $H$ parity is assigned to all 4D fields.  
The selection rule  associated with the $H$ parity conservation is useful in analyzing
production of KK excited states, higher order corrections, and so on.

The organization of the paper is the following.
In the next section the $SO(5) \times U(1)$ gauge-Higgs unification model is given 
and specified.  In Section 3 we explain how parameters of the model relevant 
for low energy physics are determined.  In Section 4 the effective potential 
$V_\eff (\theta_H)$ is re-evaluated, and $m_H$ is determined as a function of
the warp factor $z_L$.  In Section 5 a proof is given for the enhanced gauge 
invariance which in turn implies that physics is periodic in $\theta_H$ with a period 
$\pi$ in the model.  In Section 6 we show how the $H$ parity is assigned to all
4D fields.  It is shown that the action including brane interactions is invariant
under $H$ parity.  A summary is given in Section 7.

\section{Model}

The $SO(5) \times U(1)$ scheme was first proposed by Agashe, Contino, and
Pomarol,\cite{ACP}  and has been elaborated since then.
The current model is given in ref.\  \cite{HOOS} and elaborated to incorporate
leptons in ref.\ \cite{HNU}. 
It is defined in the Randall-Sundrum  (RS) warped spacetime
with a metric  
\bea
    ds^2 = G_{MN}dx^M dx^N
  = e^{-2\sigma(y)}\eta_{\mu\nu} dx^\mu dx^\nu + dy^2 , \qquad
      \label{metric1}
\eea
where $\eta_{\mu\nu} =\textrm{diag}(-1,1,1,1)$,
$\sigma(y)=\sigma(y+2L) = \sigma(-y)$, and
$\sigma(y)=k|y|$ for $|y|\leq L$.
The Planck and TeV branes are located at $y=0$ and $y=L$, respectively.  
The bulk region $0 < y < L$ is   anti-de Sitter (AdS) spacetime with a 
cosmological constant  $\Lambda = - 6k^2$.  
The warp factor $z_L \equiv e^{kL} \gg 1$  plays an important role in 
subsequent discussions.
The Kaluza-Klein (KK) mass scale is given by
\beeq
m_\KK = \frac{\pi k}{z_L -1} \sim \pi k z_L^{-1} ~.
\label{KKscale}
\eneq

The model consists of $SO(5) \times U(1)_X$ gauge fields $(A_M, B_M)$,  
bulk fermions $\Psi_a$, brane fermions $\hat \chi_{\alpha R}$, and brane scalar $\Phi$.  
\ignore{The $SO(5)$ gauge symmetry is broken to
$SO(4) \simeq SU(2)_L \times SU(2)_R$ by the orbifold boundary conditions.}
The action integral consists of the bulk and brane parts;   $S= S_\bulk + S_\brane$.
The bulk part is given by 
\beqn
&&\hskip -1cm
S_\bulk
= \int d^5x \sqrt{-G} \bigg[ -\tr \Big( {1\over 4} F^{(A)MN} F_{MN}^{(A)}
            +{1\over 2 \xi}   (f_{\textrm{\scriptsize gf}}^{(A)})^2 + 
  {\cal L}_{\textrm{\scriptsize gh}}^{(A)} \Big)  \cr
 \noalign{\kern 5pt}
&&\hskip 0.5 cm
- \Big( {1\over 4} F^{(B)MN}F_{MN}^{(B)}  
+{1\over 2\xi} (f_{\textrm{\scriptsize gf}}^{(B)})^2
     +{\cal L}_{\textrm{\scriptsize gh}}^{(B)} \Big) 
+      \sum_a  i\bar{\Psi}_a {\cal D}(c_a) \Psi_a   \bigg]      ,  \cr
\noalign{\kern 10pt}
&&\hskip -1.cm
{\cal D}(c_a) =   \Gamma^A {e_A}^M
\big( \partial_M +{1\over 8}\omega_{MBC}  [\Gamma^B, \Gamma^C]
  -ig_A A_M  -ig_B Q_{Xa} B_M \big)   -c_a \sigma'(y) ~ .
\label{action1}
\eeqn
The gauge fixing and ghost terms are
denoted as functionals with subscripts gf and gh, respectively.  
$F_{MN}^{(A)} =
 \partial_M A_N -\partial_N A_M -ig_A  [A_M, A_N]$ 
and 
$F_{MN}^{(B)} =\partial_M B_N -\partial_N B_M$. 
The $SO(5)$ gauge fields $A_M$ are decomposed as 
\beeq
 A_M = \sum_{I=1}^{10}A^I_M T^I 
 = \sum_{a_L=1}^3 A^{a_L}_M T^{a_L}+\sum_{a_R =1}^3 A^{a_R}_M T^{a_R}
 +\sum_{\hat{a}=1}^4 A^{\hat{a}}_M T^{\hat{a}} ~, 
\label{vecA1}
\eneq
where $T^{a_L, a_R}$ ($a_L, a_R=1,2,3$) and $T^{\hat{a}}$ 
($\hat{a}=1,2,3,4$) are the generators of $SO(4)\simeq SU(2)_L \times SU(2)_R$ 
and $SO(5)/SO(4)$, respectively. 

In the fermion part
$\bar{\Psi} =i\Psi^\dag \Gamma^0$
and $\Gamma^\mu$ matrices are given by
\beeq
\Gamma^\mu = 
\begin{pmatrix} & \sigma^\mu \cr \bar{\sigma}^\mu & 
\end{pmatrix}  , \quad
\Gamma^5 =\begin{pmatrix} 1 & \cr & -1  \end{pmatrix}  ,
\quad \sigma^\mu= (1,\vec{\sigma}) , \quad
 \bar{\sigma}^\mu=(-1,\vec{\sigma}) .
\label{matrix1}
\eneq
All of the bulk fermions are introduced in the vector ({\bf 5}) representation of $SO(5)$. 
The $c_a$ term in Eq.~(\ref{action1}) gives
a bulk kink mass, where
$\sigma'(y)=k\epsilon(y)$ is a periodic
step function with a magnitude $k$.
The dimensionless parameter $c_a$ plays an important role controlling profiles
of fermion wave functions.

The orbifold boundary conditions at $y_0=0$ and $y_1=L$  are given by
\beqn
&&\hskip -1.cm
\begin{pmatrix}    A_\mu \cr A_y    \end{pmatrix}  (x,y_j-y)
= P_j  \begin{pmatrix}    A_\mu \cr - A_y  \end{pmatrix}  (x,y_j+y) P_j^{-1} , \cr
\noalign{\kern 10pt}
&&\hskip -1.cm
\begin{pmatrix}    B_\mu \cr B_y    \end{pmatrix}  (x,y_j-y)
= \begin{pmatrix}    B_\mu \cr - B_y  \end{pmatrix}  (x,y_j+y) , \cr
\noalign{\kern 10pt}
&&\hskip -1.cm
\Psi_a (x,y_j-y) = P_j \Gamma^5 \Psi_a (x,y_j+y) , \cr
\noalign{\kern 10pt}
&&\hskip -1.cm
P_j =\textrm{diag} \, (-1,-1,-1,-1,+1)~.
\label{BC1}
\eeqn
The $SO(5)\times U(1)_X$ symmetry is reduced to
$SO(4)\times U(1)_X \simeq SU(2)_L \times SU(2)_R \times U(1)_X$
by the orbifold boundary conditions.
Rigorously speaking, various orbifold boundary conditions fall into a finite number of 
equivalence classes of boundary conditions.\cite{YH2, HHHK, HHK}  
In each class apparently different boundary
conditions are related to each other by  Wilson line phases.
The physical symmetry of the true vacuum in each equivalence class of boundary
conditions is determined at the quantum level.

The 4D Higgs field, which is a  doublet both in $SU(2)_L$ and in $SU(2)_R$,
appears as a zero mode in the $SO(5)/SO(4)$ part of the fifth  dimensional component 
of the vector potential $A_y^{\hat a} (x,y)$.  
Without loss of generality one assumes  $\la A_y^{\hat a} \ra  \propto \delta^{a4}$ 
when the EW symmetry is spontaneously broken. 
The generator $T^{\hat 4}$ is given by 
$(T^{\hat 4})_{ab} =( i /\sqrt{2}) (\delta_{a5} \delta_{b4} -\delta_{a4} \delta_{b5}) $
in the vectorial representation, 
whereas $T^{\hat 4} = ( 1 /2\sqrt{2}) I_2 \otimes \tau_1$ in the spinorial representation.
The Wilson line phase $\theta_H$ is given by 
\beeq
\exp \Big\{\frac{i}{2}\theta_H  \cdot 2\sqrt2 \, T^{\hat{4}}  \Big\}
=\exp \bigg\{ ig_A \int^{L}_{0} dy  \la A_y \ra \bigg\}
\label{Higgs1}
\eneq
so that the 4D neutral Higgs field $H(x)$ appears as \cite{HK}
\beqn
&&\hskip -1.cm
A^{\hat{4}}_y (x,y)=\big\{ \theta_H f_H + H(x) \big\} u_H(y) +\cdots ~, \cr
\noalign{\kern 10pt}
&&\hskip -1.cm
f_H = \frac{2}{g_A} \sqrt{\frac{k}{z_L^2 -1}}
~,~~ u_H(y) = \sqrt{\myfrac{2k}{z^2_L-1}} \, e^{2ky} ~~~
(0\le y \le L)~.
\label{Higgs2}
\eeqn

For each generation two vector multiplets $\Psi_1$ and $\Psi_2$ for quarks
and two vector multiplets $\Psi_3$ and $\Psi_4$ for leptons are introduced.
Each vector multiplet, $\Psi$, is decomposed into one $(\onehalf, \onehalf)$, 
$\check \Psi$,  and one $(0,0)$ of $SU(2)_L \times SU(2)_R$.  
We denote $\Psi_a$'s , for the third generation, as
\beqn
&&\hskip -1.cm
\Psi_1= \big(  \check \Psi_1 , ~ t'  ~\big)_{2/3} ~,~~
 \check \Psi_1 = \begin{pmatrix} T & t \cr B & b \end{pmatrix} 
 \equiv \bigg( Q_1  ,    q  \bigg) ~,  \cr
\noalign{\kern 10pt}
&&\hskip -1.cm
\Psi_2= \big(  \check \Psi_2 , ~ b'  ~\big)_{-1/3} ~, ~~ 
\check \Psi_2 =  \begin{pmatrix} U &X \cr D & Y \end{pmatrix} 
\equiv \bigg( Q_2  ,  Q_3 \bigg) ~, \cr
\noalign{\kern 10pt}
&&\hskip -1.cm
\Psi_3= \big(  \check \Psi_3 , ~ \tau'  ~\big)_{-1} ~,  ~~
\check \Psi_3 = \begin{pmatrix} \nu_\tau & L_{1X} \cr  \tau  &  L_{1Y} \end{pmatrix}
\equiv \bigg( \ell ,  L_1 \bigg) ~, \cr
\noalign{\kern 10pt}
&&\hskip -1.cm
\Psi_4= \big(  \check \Psi_4 , ~ \nu_\tau '  ~\big)_{0} ~, ~~ 
\check \Psi_4 = \begin{pmatrix}  L_{2X} & L_{3X} \cr L_{2Y}  & L_{3Y} \end{pmatrix}
\equiv \bigg( L_2  ,   L_3 \bigg) ~.
\label{bulkF1}
\eeqn
Subscripts $2/3$ etc. represent $U(1)_X$ charges, $Q_X$, of  $\Psi_a$'s.
$q$, $Q_j$, $\ell$, and $L_j$ are $SU(2)_L$ doublets.
The electromagnetic charge $Q_\EM$ is given, a posteriori, by
\beeq
Q_\EM = T^{3_L} + T^{3_R} + Q_X ~.
\label{charge1}
\eneq
Each $\Psi_a$ has its bulk mass parameter $c_a$.  Consistent results are
obtained by taking $c_1 = c_2 \equiv c_q$ and $c_3=c_4 \equiv c_\ell$ 
for each generation.

The additional brane fields are introduced on the Planck brane at $y=0$.  
The  brane scalar $\Phi$ belongs to $(0,{1\over 2})$ of $SU(2)_L \times SU(2)_R$
with $Q_X = - \onehalf$, whereas the right-handed
brane fermions $\hat \chi_{\alpha R}^q$ and $\hat \chi_{\alpha R}^\ell$ belong to
 $({1\over 2}, 0)$.  The brane fermions are
 \beqn
 &&\hskip -1.cm
 \hat \chi_{1R}^q = \begin{pmatrix} \hat T_R \cr \hat B_R \end{pmatrix}_{7/6} , ~~
 \hat \chi_{2R}^q = \begin{pmatrix} \hat U_R \cr \hat D_R \end{pmatrix}_{1/6}, ~~
 \hat \chi_{3R}^q = \begin{pmatrix} \hat X_R \cr \hat Y_R \end{pmatrix}_{-5/6} ,   \cr
 \noalign{\kern 10pt}
&&\hskip -1.cm
 \hat \chi_{1R}^\ell = \begin{pmatrix} \hat L_{1XR} \cr \hat L_{1YR} \end{pmatrix}_{-3/2} , ~~ 
 \hat \chi_{2R}^\ell = \begin{pmatrix} \hat L_{2XR} \cr \hat L_{2YR} \end{pmatrix}_{1/2} , ~~ 
 \hat \chi_{3R}^\ell = \begin{pmatrix} \hat L_{3XR} \cr \hat L_{3YR} \end{pmatrix}_{-1/2} .
 \label{braneF1}
 \eeqn
Subscripts $7/6$ etc. represent $Q_X$ charges of  $\hat \chi_R$'s.
The brane part of the action is given by
\beqn
&&\hskip -1.cm
S_\brane
= \int d^5 x  \sqrt{-G} ~   \delta(y) \bigg\{
   -(D_\mu \Phi)^\dag D^\mu \Phi  -\lambda_\Phi  (\Phi^\dag \Phi -w^2)^2  \cr
\noalign{\kern 5pt}
&&\hskip 4. cm
+ \sum_{\alpha=1}^3 \Big( \hat{\chi}_{\alpha R}^{q \dag}
  \,  i \bar{\sigma}^\mu   D_\mu \hat{\chi}_{\alpha R}^q
   +  \hat{\chi}_{\alpha R}^{\ell \dag}
\,  i \bar{\sigma}^\mu   D_\mu \hat{\chi}_{\alpha R}^\ell  \Big)     \cr
\noalign{\kern 5pt}
&&\hskip -.5cm
-   i\Big[ \kappa_1^q \, \hat{\chi}_{1 R}^{q \dag} \check \Psi_{1L} \tilde \Phi 
  + \tilde \kappa^q \,  \hat{\chi}_{2 R}^{q \dag} \check \Psi_{1L}  \Phi
  + \kappa_2^q \,  \hat{\chi}_{2 R}^{q \dag} \check \Psi_{2L} \tilde \Phi 
  + \kappa_3^q \,  \hat{\chi}_{3 R}^{q \dag} \check \Psi_{2L} \Phi 
  - (\hbox{h.c.}) \Big] \cr
\noalign{\kern 10pt}
&&\hskip -.5cm
-   i\Big[ \tilde \kappa^\ell \, \hat{\chi}_{3 R}^{\ell \dag} \check \Psi_{3L} \tilde \Phi 
  + \kappa_1^\ell \,  \hat{\chi}_{1 R}^{\ell \dag} \check \Psi_{3L}  \Phi
  + \kappa_2^\ell \,  \hat{\chi}_{2 R}^{\ell \dag} \check \Psi_{4L} \tilde \Phi 
  + \kappa_3^\ell \,  \hat{\chi}_{3 R}^{\ell \dag} \check \Psi_{4L} \Phi 
  - (\hbox{h.c.}) \Big] \bigg\} ~, \cr
\noalign{\kern 5pt}
&&\hskip -0.cm
D_\mu \Phi = \Big( \dd_\mu - i   g_A \sum_{a_R=1}^3  A_\mu^{a_R} T^{a_R}
    + i  {1\over 2}g_B B_\mu \Big) \Phi  ~, 
     ~~ \tilde \Phi = i \sigma_2 \Phi^* ~,  \cr
\noalign{\kern 0pt}
&&\hskip -0.cm
D_\mu \hat \chi = \Big( \dd_\mu  - i  g_A \sum_{a_L=1}^3  A_\mu^{a_L} T^{a_L}
      -i  Q_X g_B B_\mu \Big) \hat \chi  ~.
\label{action2}
\eeqn
The  action $S_\brane$ is manifestly invariant under 
$SU(2)_L \times SU(2)_R \times U(1)_X$.  
The Yukawa couplings above exhaust all possible ones preserving the symmetry.

The non-vanishing vev $w$ have two important consequences.  
We need to assume only that $w \gg m_\KK$.  
Firstly the $SU(2)_R \times U(1)_X$ symmetry is spontaneously broken down to $U(1)_Y$ 
and the zero modes of four-dimensional gauge fields of 
$SU(2)_R \times U(1)_X$ become massive except for the $U(1)_Y$ part.  
They acquire masses of $O(m_\KK)$ as a result of the effective change of 
boundary conditions for low-lying modes in the Kaluza-Klein towers.
Secondly the non-vanishing vev $w$ induces mass couplings between
brane fermions and bulk fermions;
\beqn
&&\hskip -1.cm
S_\brane^{\rm mass} = \int d^5 x  \sqrt{-G} ~   \delta(y) \bigg\{
-\sum_{\alpha=1}^3  
     i \mu_\alpha^q  (\hat{\chi}_{\alpha R}^{q\dag} Q_{\alpha L}
    -Q_{\alpha L}^\dag \hat{\chi}_{\alpha R}^q) 
    -i \tilde{\mu}^q (\hat{\chi}_{2R}^{q \dag} q_L   - q_L^\dag \hat{\chi}_{2R}^q) \cr
\noalign{\kern 5pt}
&&\hskip 3.cm
- \sum_{\alpha=1}^3  
    i \mu_\alpha^\ell  (\hat{\chi}_{\alpha R}^{\ell \dag} L_{\alpha L}
    -L_{\alpha L}^\dag \hat{\chi}_{\alpha R}^\ell ) 
    -i \tilde{\mu}^\ell  (\hat{\chi}_{3R}^{\ell \dag} \ell_L   
    -\ell_L^\dag \hat{\chi}_{3R}^\ell)     \bigg\} ~,  \cr
\noalign{\kern 10pt}
&&\hskip 2.cm
\frac{\mu_\alpha^q}{\kappa_\alpha^q} = \frac{\tilde \mu^q}{\tilde \kappa^q}
= \frac{\mu_\alpha^\ell}{\kappa_\alpha^\ell} = \frac{\tilde \mu^\ell}{\tilde \kappa^\ell}
=  w ~,
\label{action3}
\eeqn
Assuming that all $\mu^2 \gg m_\KK$,  all of the exotic zero modes of
the bulk fermions acquire large masses of $O(m_\KK)$.  
It has been shown that all of the 4D anomalies associated with 
$SU(2)_L \times SU(2)_R \times U(1)_X$ gauge symmetry are cancelled.\cite{HNU}
The $SU(2)_L \times U(1)_Y$  is further broken down to $U(1)_\EM$ 
by the Hosotani mechanism.
The spectrum of the resultant light particles are the same as  in the standard model.

\section{Parameters of the model \label{sec:parameters}}

The parameters of the model relevant for low energy physics are 
$k$, $z_L=e^{kL}$, $g_A$, $g_B$,  the bulk mass parameters 
$(c_q, c_\ell)$ and the brane mass ratios 
$(\tilde \mu^q / \mu^q_2, \tilde \mu^\ell / \mu^\ell_3)$.  
All other parameters are irrelevant at low energies, provided that
$w$,  $\mu^2$'s are much larger than $m_\KK$.
The value of $\theta_H$ is determined dynamically to be $\pm \onehalf \pi$
as shown in Section 4, where the electroweak symmetry is broken 
and $W$, $Z$, quarks and leptons acquire non-vanishing masses.  

All parameters are fixed at $\theta_H = \pm \onehalf \pi$.  
Three of the four parameters $k$, $z_L=e^{kL}$, $g_A$, $g_B$ are determined
from the $Z$ boson mass $m_Z$, the weak gauge coupling $g_w$, and 
the Weinberg angle $\sin^2 \theta_W$.  The one parameter, say, $z_L$ remains
undetermined.   
In the fermion sector let us, for the moment,  forget about
the mixing among generation and consider quark and lepton
masses in each generation separately.  Take the first generation as an example.
In the quark sector the bulk mass $c_q$ and the ratio $\tilde \mu^q / \mu^q_2$
are determined from $m_u$ and $m_d$.   Similarly in the lepton sector
$c_\ell$ and $ \tilde \mu^\ell / \mu^\ell_3$ are determined from 
$m_e$ and $m_{\nu_e}$.  As $m_{\nu_e} \ll m_e$,  all of the results 
discussed below do not depend on the unknown value of $m_{\nu_e}$.
If neutrinos were massless, one could delete $\Psi_4$, $\hat \chi^\ell_{2R}$, 
$\hat \chi^\ell_{3R}$,  and all of the associated couplings from the model.
In this case $m_e$ determines $c_\ell$ in the first generation.
The generation mixing can be incorporated by considering 3-by-3 matrices
for the brane masses $\mu$'s,  the investigation of which is reserved 
for future.

Once the value of $z_L$ is specified, all the relevant parameters of the model 
are determined.  The spectra of particles and their KK towers, 
their wave functions in the fifth dimension, and all interaction couplings can 
be calculated.
The effective potential for $\theta_H$ is evaluated at the one loop level, 
from which the mass of the 4D Higgs boson, $m_H$, is predicted.
It will be found that $m_H$ is about $70 \sim 135\,$GeV for
$z_L = 10^{5} \sim 10^{15}$.  
Conversely the remaining one parameter
$z_L$ is fixed, once the Higgs boson mass $m_H$ is given.

As typical reference values we take  the warp factors $z_L = 10^5, 10^{10}, 10^{15}$.
The values in Table~\ref{tab:input} are taken, as input parameters, 
for the masses of quarks, leptons and  gauge boson. 
The masses of quarks and charged leptons except for 
$t$ quark are quoted from Ref.~\cite{Xing:2007fb}.
The masses of $Z$ boson and $t$ quark are
the central values in the Particle Data Group review~\cite{Amsler:2008zzb}.
The couplings $\alpha$ and $\alpha_s$ are 
also quoted from Ref.~\cite{Amsler:2008zzb}.
In the present analysis, the neutrino masses have negligible effects.

The remaining parameter,  $\sin^2 \theta_W$, needs to be determined
by global fit.  We choose $\sin^2 \theta_W =0.2312, 0.2285$ for 
$z_L = 10^{15}, 10^5$, respectively.   
Since complete one-loop analysis is not available in the 
gauge-Higgs unification scenario at the moment,  there remains
ambiguity in the value of $\sin^2 \theta_W$.

\begin{table}[htb]
\begin{center}
\caption{Input parameters for the masses and couplings of the model.
The masses are  in an unit of GeV.   All masses except for $m_t$ are
at the $m_Z$ scale.}
\label{tab:input}
\vskip 10pt
\begin{tabular}{|ccccccc|}
\hline 
$m_Z$ & $m_u$ & $m_d$ & $m_s$
& $m_c$ & $m_b$ & $m_t$ \\
\hline
91.1876 & 1.27 $\times 10^{-3}$ &
2.90 $\times 10^{-3}$ &
0.055 & 0.619 & 2.89 & 171.17 \\
\hline
\noalign{\kern 5pt}
\hline 
\multicolumn{2}{|c}{$m_e$} & 
\multicolumn{2}{c}{$m_\mu$} & 
$m_\tau$ & $\alpha(m_W)$ & $\alpha_s (m_Z)$ \\
\hline
\multicolumn{2}{|c}{$0.486570161 \times 10^{-3}$} &
\multicolumn{2}{c}{$102.7181359 \times 10^{-3}$} &
1.74624 & 1/128 & 0.1176 \\
\hline
\end{tabular}
\end{center}
\end{table}

\section{EW symmetry breaking and the Higgs boson mass}

After the spontaneous breaking of $SU(2)_R \times U(1)_X$ to $U(1)_Y$
the model has the standard model  (SM) symmetry $SU(2)_L \times U(1)_Y$.  
The SM symmetry is dynamically broken down to $U(1)_\EM$ by the Hosotani
mechanism.  To confirm it,  one need to evaluate the effective potential
$V_\eff (\theta_H)$ for the Wilson line phase, $\theta_H$.

The effective potential $V_\eff (\theta_H)$ has been evaluated in ref.\ \cite{HOOS}.
The model has one free parameter, $z_L$, to be fixed.   It is shown below that
$V_\eff (\theta_H)$ is minimized at $\theta_H = \pm \onehalf \pi$
provided $z_L > z_L^c$.  The Higgs boson mass $m_H$ is determined from 
the curvature of $V_\eff (\theta_H)$ at the minimum.
This effective potential $V_\eff $  is important  in discussing the radion 
stabilization as well.\cite{Sakamura2010a, Sakamura2010b}

The effective potential at the one loop level is determined by the
spectrum of the particles.  Suppose that the spectrum of a given particle, 
$m_n (\theta_H) = k \lambda_n(\theta_H)$,
is determined by roots of an equation $1 + \tilde Q (\lambda_n; \theta_H) = 0$.
Then \cite{Oda1, Falkowski1}
\beqn
&&\hskip -1.cm
V_\eff (\theta_H) = \sum_{\rm particles}
\pm \frac{1}{2} \int \frac{d^4 p}{(2\pi)^4} 
\sum_n \ln \big( p^2 + m_n(\theta_H)^2  \big)  \cr
\noalign{\kern 5pt}
&&\hskip 0.5 cm
= \sum_{\rm particles}
\pm I [Q(q; \theta_H)] ~, \cr
\noalign{\kern 10pt}
&&\hskip -1cm
 I [ Q(q; \theta_H) ] =  \frac{(k z_L^{-1})^4}{(4\pi)^2} \int_0^\infty dq  \, q^3 
\ln \Big\{ 1 + Q(q; \theta_H) \Big\} ~,  \cr
\noalign{\kern 10pt}
&&\hskip - 1.cm
Q(q; \theta_H ) = \tilde Q (i q z_L^{-1} ; \theta_H) ~.
\label{effV1}
\eeqn
Here $\pm$ corresponds to bosons or fermions.  
The sums extend over all degrees of particle freedom.
The $\theta_H$-dependent part of $V_\eff (\theta_H)$ is known to 
be finite.\cite{YH1, Hosotani2005}
The integral over $q$ is saturated in the range $0 < q < 10$.

\begin{figure}[t,b]
\centering  \leavevmode
\includegraphics[height=4.3cm]{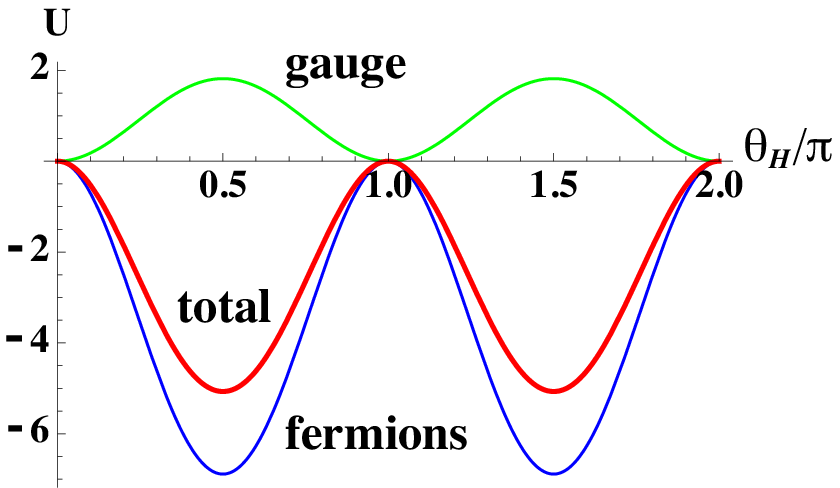}
\hskip 5pt
\includegraphics[height=4.3cm]{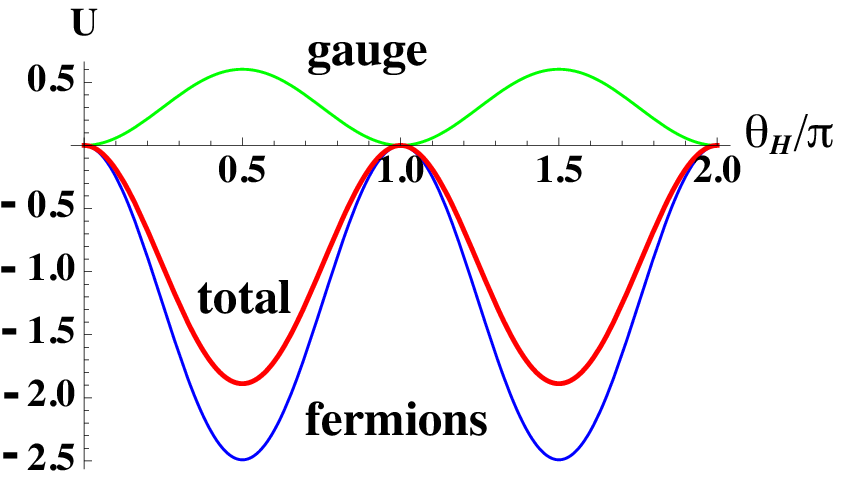}
\caption{The effective potential  $V_\eff (\theta_H)$ in the model. The plot is for  
$U (\theta_H/\pi) = (4\pi)^2 (kz_L^{-1})^{-4} \, V_\eff$ 
at  $z_L = 10^{5}$ (left) and $z_L = 10^{15}$ (right).  
Green, blue, and red curves represent $V_\eff^{\rm gauge}$, $V_\eff^{\rm fermion}$,
and $V_\eff$, respectively. 
The global minima are located at 
$\theta_H = \onehalf \pi $ and $\frac{3}{2}\pi$, where  the EW symmetry  
dynamically breaks down to $U(1)_{\EM}$.}
\label{effV-total-fig}
\end{figure}

It is convenient to introduce 
\beqn
&&\hskip -1cm 
Q_0(q;  c,  \theta_H) = \frac{z_L}{q^2} 
\frac{\sin^2 \theta_H}{\hat F_{c - \onehalf,c - \onehalf} (qz_L^{-1}, q) 
              \hat F_{c + \onehalf, c + \onehalf} (qz_L^{-1}, q) }  ~~, \cr
\noalign{\kern 10pt}
&&\hskip -1cm 
\hat F_{\alpha ,\beta} (u, v) = I_\alpha (u) K_\beta(v) 
- e^{- i (\alpha - \beta) \pi}  K_\alpha (u) I_\beta(v)  ~~,
\label{function1}
\eeqn
where $I_\alpha$ and $K_\alpha$ are modified Bessel functions.
The contributions of gauge fields to $V_\eff (\theta_H)$ are given by
\beeq
V_\eff (\theta_H)^{\rm gauge} = 
4 I [ \onehalf Q_0(q;  \onehalf, \theta_H)  ] 
+ 2   I \Big[ \frac{1}{2 \cos^2 \theta_W} Q_0(q;  \onehalf, \theta_H)  \Big] 
+ 3  I [ Q_0(q;  \onehalf, \theta_H) ] ~, 
\label{effV2}
\eneq
whereas contributions of fermions are given by\footnote{The color factor 3 
was missing for the contributions of quarks  in ref.\ \cite{HOOS}.
The authors thank T.\ Ohnuma and Y.\ Sakamura for pointing out this error.}
\beqn
&&\hskip -1cm
V_\eff (\theta_H)^{\rm fermion} \cr
\noalign{\kern 10pt}
&&\hskip -1.cm
\simeq - 12  \sum_{\rm quarks} \bigg\{ 
I \Big[ \frac{1}{2(1+r_q)} Q_0 (q;  c_q, \theta_H) \Big] 
+ I \Big[ \frac{r_q}{2(1+r_q)} Q_0 (q; c_q, \theta_H) \Big] \bigg\}  \cr
\noalign{\kern 10pt}
&&\hskip -0.5 cm
- 4  \sum_{\rm leptons} \bigg\{ 
I \Big[ \frac{1}{2(1+r_\ell)} Q_0 (q;  c_\ell, \theta_H) \Big] 
+ I \Big[ \frac{r_\ell}{2(1+r_\ell)} Q_0 (q; c_\ell, \theta_H) \Big] \bigg\} ~, \cr
\noalign{\kern 10pt}
&&\hskip 1cm
r_q = \frac{(\tilde \mu^q)^2}{(\mu_2^q)^2}  ~~,~~
r_\ell = \frac{(\mu_3^\ell)^2}{(\tilde \mu^\ell)^2}~~.
\label{effV3}
\eeqn
In $V_\eff^{\rm fermion}$ each integral $I$ sensitively depends
on the value of the bulk mass parameter $c_q$ or $c_\ell$. 
Contributions from fermion multiplets with $c > 0.6$ are negligible compared with
$V_\eff^{\rm gauge}$.  The  relevant contribution comes solely from 
the multiplet containing a top quark.  
The top quark contribution dominates over $V_\eff^{\rm gauge}$ in the RS warped space, 
yielding the minima of $V_\eff$ at $\theta_H = \pm \onehalf \pi$.  
In fig.\ \ref{effV-total-fig},  $V_\eff (\theta_H)$ is displayed for $z_L = 10^5$
and $10^{15}$.  
Contributions from light quarks and leptons are
suppressed by a factor of $\sim 10^6$.   
The top quark  dominates over gauge fields for $z_L = 10^{15}$ more than 
for $z_L = 10^{5}$.

We observe that
\beeq
V_\eff (\theta_H + \pi) = V_\eff (\theta_H) = V_\eff (- \theta_H) ~.
\label{effV4}
\eneq
It is important in the first equality that all bulk fermions are introduced
in the vector representation of $SO(5)$. 
If there were a bulk fermion, say, in the spinor representation of $SO(5)$,
the $\theta_H$-dependence in $I$ in (\ref{effV3})  would contain 
$\sin^2 \onehalf \theta_H$ instead of $\sin^2  \theta_H$.  
If all bulk fermions were in the spinor representation, the minimum of
$V_\eff$ would be located either at $\theta=0$ or $\pi$ so that 
the EW symmetry would be unbroken.  

We also remark that the scale of the depth of the effective potential is
given by $m_\KK / (2 \pi^{3/2})$.  As the universe  expands and cools down,
the electroweak symmetry breaking is expected to take place at a 
temperature of the electroweak scale.  To determine the precise value
one needs to evaluate the effective potential at finite temperature.

The mass of the 4D neutral Higgs boson is determined from the curvature of 
the effective potential at the minimum.    Making use of (\ref{Higgs2}), one finds 
\beeq
m_H^2 = \frac{1}{f_H^2} \frac{d^2 V_\eff}{d \theta_H^2} 
\bigg|_{\theta_H = \onehalf\pi} ~.
\label{Higgs3}
\eneq
It follows from (\ref{effV3}) that 
\beqn
&&\hskip -1.cm
m_H^2 \simeq
\frac{g_w^2 kL m_\KK^2}{64 \pi^4} 
\bigg\{ - 4 G[\onehalf \bar Q_0(q, \onehalf)] 
-  2 G\Big[ \frac{1}{2 \cos^2 \theta_W} \bar Q_0(q, \onehalf) \Big] 
- 3  G[\bar Q_0(q, \onehalf)]  \cr
\noalign{\kern 10pt}
&&\hskip 1.5cm
+ 12  \sum_{\rm quarks} \bigg(   G \Big[ \frac{1}{2(1+r_q)} \bar Q_0(q,c_q) \Big] 
+  G \Big[ \frac{r_q}{2(1+r_q)} \bar Q_0(q,c_q) \Big] \bigg)  \cr
\noalign{\kern 10pt}
&&\hskip 1.5cm
+ 4  \sum_{\rm leptons} \bigg(   G \Big[ \frac{1}{2(1+r_\ell)} \bar Q_0(q,c_\ell) \Big] 
+  G \Big[ \frac{r_\ell}{2(1+r_\ell)} \bar Q_0(q, c_\ell) \Big] \bigg) \bigg\} ~,  \cr
\noalign{\kern 10pt}
&&\hskip -1.cm
G[f(q)] = \int_0^\infty dq \, q^3 \frac{2 f(q)}{1 + f(q)}  ~~,
~~  \bar Q_0(q, c ) \equiv Q_0(q;  c , \onehalf \pi) ~~.
\eeqn
Among fermion multiplets, only the top quark multiplet gives an appreciable 
contribution.  The result is summarized in Table \ref{mH-table}.

Higgs bosons become stable in the model.  They can become the dark matter in the 
universe.  It was shown in ref.\ \cite{HKT} that the mass density of the dark matter 
determined by the WMAP data is reproduced with $m_H \sim 70\,$GeV.
This value of $m_H$ is obtained with $z_L \sim 10^5$ in the current model.

\begin{table}[b,t]
\begin{center}
\begin{tabular}{|c|c||c|c|c|c|c|} 
\noalign{\kern 15pt}
\hline
$z_L= e^{kL}$ & $\sin^2 \theta_W$ & $k$(GeV) & $m_\KK$(GeV) 
   & $c_{\rm top}$ & $m_H$(GeV)   & $m_W^{\rm tree}$(GeV)\\ \hline 
$10^{15}$  &0.2312    &~$4.666 \times 10^{17}$~ 
    & 1,466 & $\,$ 0.432 $\,$ & ~ 135 ~    &79.82  \\ \hline
$10^{10}$  & 0.23 & $3.799 \times 10^{12}$ & 1,194 & 0.396 
    & ~ 108 ~   &79.82\\ \hline
~$10^{5}$~   &0.2285  & $2.662 \times 10^{7}$~ & ~~836 & 0.268 
   & ~ ~72 ~  &79.70 \\ \hline
\end{tabular}
\end{center}
\caption{The Higgs boson mass $m_H$.  
Relevant input parameters are $m_Z = 91.1876\,$GeV, 
$\alpha_w = 1/128$  and $m_t = 171.17 \,$GeV.
The AdS curvature $k$ and $W$ mass at the tree level are also listed.
}
\label{mH-table}
\end{table}

\begin{figure}[t,b]
\centering  \leavevmode
\includegraphics[height=4.3cm]{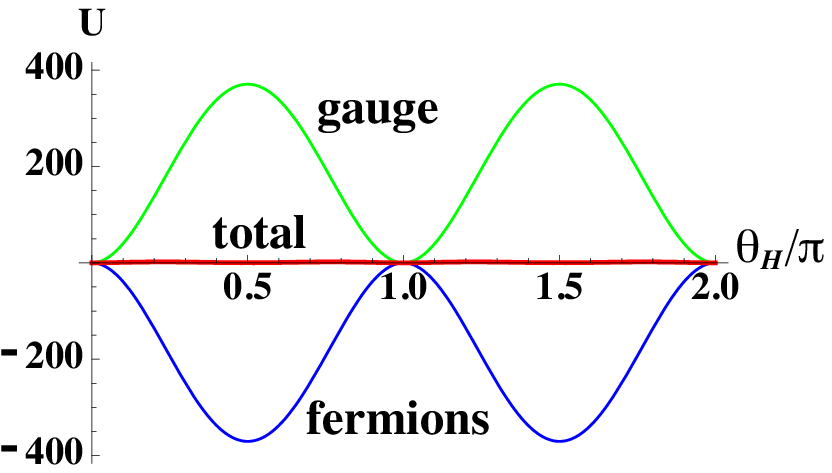}
\hskip 5pt
\includegraphics[height=4.3cm]{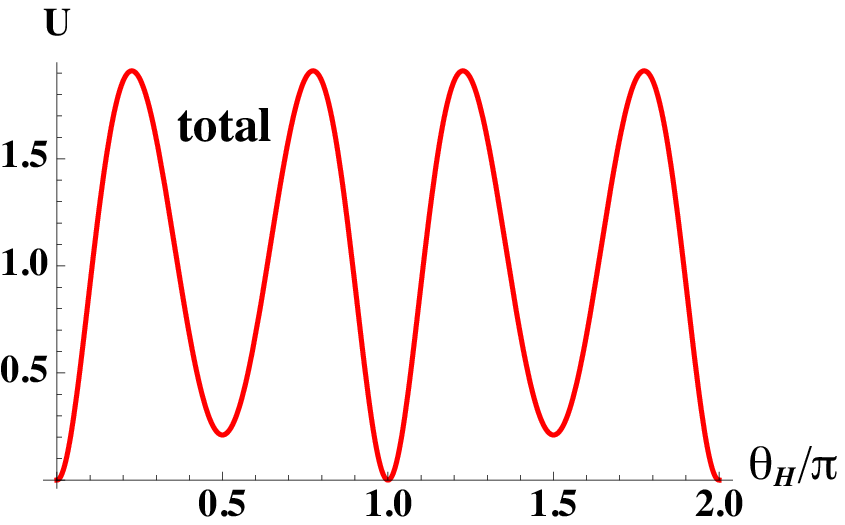}
\caption{The critical behavior near $z_L =1.67$, 
below which $V_\eff$ is minimized at $\theta_H = 0, \pi$.
}
\label{critical-fig}
\end{figure}

It is curious to examine whether or not the EW symmetry is 
broken in the flat spacetime limit.   As shown in ref.\ \cite{HOOS}
the top quark mass $m_t \sim 170\,$GeV cannot be realized
for $z_L < 900$.   It is possible to consider the flat spacetime limit
($k \go 0, ~z_L \go 1$) by taking  the bulk mass $c=0$ for the top
quark multiplet.  It is found that around $z_L^c \sim 1.67$ the phase transition
takes place.  The transition is weakly first-order.  Below $z_L$ the global
minima of $V_\eff$ are located at $\theta_H = 0, \pi$ where the EW 
symmetry remains unbroken.  See fig.\ \ref{critical-fig}.

\section{Enhanced gauge invariance}

In this section we show that the theory is invariant under the shift 
$\theta_H \go \theta_H + \pi$ to all order in perturbation theory.
In other words the physics is periodic in $\theta_H$ with a period  $\pi$.
This property follows from the enhanced gauge invariance in the model
in which (i) the bulk fermions are all in the vector representation of $SO(5)$,
and (ii) the brane fermions and brane scalar are introduced only on 
one of the two branes, say, on the Planck brane.

To see it we consider an $SO(5)$ gauge transformation
$A_M' = \Omega A_M \Omega^{-1} + (i/g_A) \Omega \dd_M \Omega^{-1}$
where
\beeq
\Omega(y; \alpha)  = \exp \Big\{ i \alpha q(y) \, T^{\hat 4} \Big\} ~~,~~
q(y) = g_A  f_H \int_0^y dy' \,  u_H(y') ~~.
\label{largeG1}
\eneq
$u_H(y)$ in $0 \le y \le L$ is given by (\ref{Higgs2}),  and is extended in other
regions by $u_H(-y)=u_H(y) = u_H(y+ 2L)$.  
It follows that
\beeq
q(y) + q(-y) = 0 ~~,~~ q(L+y) + q(L-y) = 2 \sqrt{2} ~.
\label{largeG2}
\eneq
In the fundamental region $0 \le y \le L$ 
\beeq
\Omega(y; \alpha) = \exp \bigg\{ i \sqrt{2} \, \alpha \,
\frac{e^{2ky} - 1}{z_L^2 - 1} \, T^{\hat 4}  \bigg\}~.
\eneq
This gauge transformation shifts the Wilson line phase $\theta_H$ to
$\theta_H' = \theta_H + \alpha$.  
The fields in the new gauge satisfy the boundary condition (\ref{BC1}) with
$P_j$ replaced by $P_j(\alpha) = \Omega(y_j - y ; \alpha) P_j \Omega(y_j + y; \alpha)^{-1}$.
Note that $P_j(\alpha)$ is independent of $y$. 

In the vectorial representation $P_j = \diag (-1,-1,-1,-1,1)$ and
$(T^{\hat 4})_{ab} = (i/\sqrt{2}) (\delta_{a5} \delta_{b4} - \delta_{a4} \delta_{b5})$
so that $P_0 (\pi)= P_0$ and $P_1 (\pi) = P_1$.  
In the spinorial representation $P_j =  I_2 \otimes \tau_3$ and
$T^{\hat 4} = (1/2\sqrt{2}) I_2 \otimes \tau_1$ so that 
$P_0 (\pi)= P_0$ and $P_1 (\pi) = - P_1$.  As $\Omega(0; \alpha) = 1$,
the brane fermions and scalar are not affected by this gauge transformation.
It follows that the model under consideration  is invariant 
under the large gauge transformation $\Omega (y ; \pi)$, that is to say, the theory
is periodic in $\theta_H$ with a period $\pi$.  
It implies, for instance, that $V_\eff (\theta_H + \pi) = V_\eff (\theta_H)$ 
to all order in perturbation theory.   The mirror reflection symmetry under $y \go -y$
leads to $V_\eff (-\theta_H) = V_\eff (\theta_H)$.  Combining these two, one finds
that $V_\eff (\theta_H)$ is symmetric around $\theta_H = \pm \onehalf \pi$
to all order in perturbation theory.  In the previous section we have observed that 
$V_\eff (\theta_H)$ is minimized at $\theta_H = \pm \onehalf \pi$ at the one-loop
level.  The location of the minimum will not be shifted in one direction by radiative
corrections.  $\theta_H = \pm \onehalf \pi$ remains as an extremum of $V_\eff$.

We stress that the above property would be lost if there were, for instance,
a bulk fermion in the spinor representation of $SO(5)$.  
Furthermore $\Omega (L; \pi) = \exp \{ i \sqrt{2} \pi T^{\hat 4} \} \not= 1$
in either vectorial or spinorial representation.  If brane fields were introduced
on the TeV brane at $y=L$ as well as on the Planck brane  at $y=0$,
then the enhanced periodicity would be lost in general.
In passing  $\Omega (L; 2\pi) = 1$ or $-1$ in the vectorial  or spinorial representation,
respectively.

\section{$H$ parity \label{sec:hparity}}

We expand all fields around the vacuum $\theta_H = \onehalf \pi$.  
It has been shown in ref.\ \cite{HKT} that  the $H$ parity ($P_H$) conservation
results among the low energy fields as a result of the enhanced gauge invariance
and the mirror reflection symmetry in the fifth dimension.
The neutral physical Higgs boson
is $P_H$ odd, whereas all other particles in the standard model  are $P_H$ even.  
It follows that the lightest $P_H$ odd particle,
the Higgs boson, is absolutely stable.  

A natural question arises as to whether all fields including KK modes can
be classified with respect to $P_H$.  We show that one can assign  definite $H$ parity
to all fields at $\theta_H = \onehalf \pi$
and that both the bulk action (\ref{action1}) and the brane action (\ref{action2}) 
are invariant under $P_H$.  
As we shall see below, $P_H$ interchanges $SU(2)_L$ and $SU(2)_R$ and flips the 
sign of $T^{\hat 4}$.   The $P_H$ symmetry is similar to the  $P_{LR}$ symmetry
discussed by Agashe, Contino, Da Rold and Pomarol  \cite{Agashe2006}, which  protects the 
$Z b \bar b$ coupling from radiative corrections.

\begin{figure}[htb]
\begin{center}
\vspace{1ex}
\includegraphics[height=4.5cm]{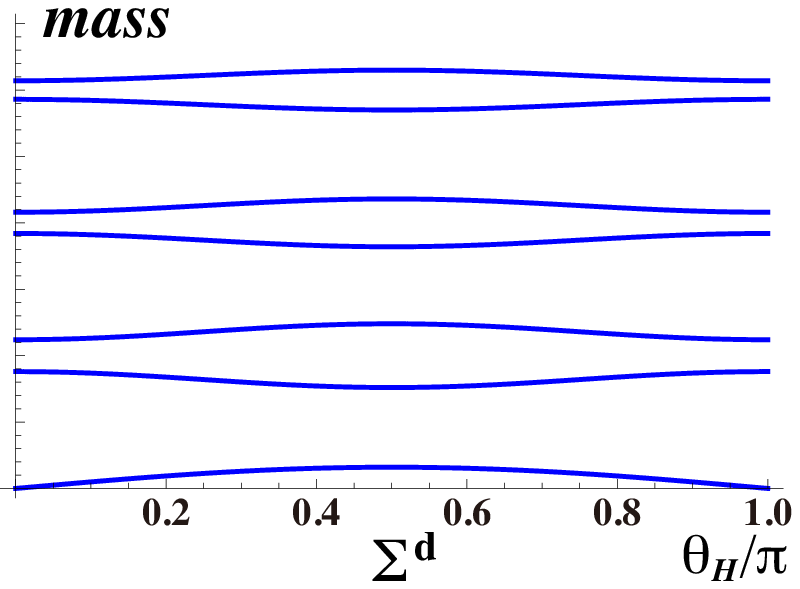}
\qquad 
\includegraphics[height=4.5cm]{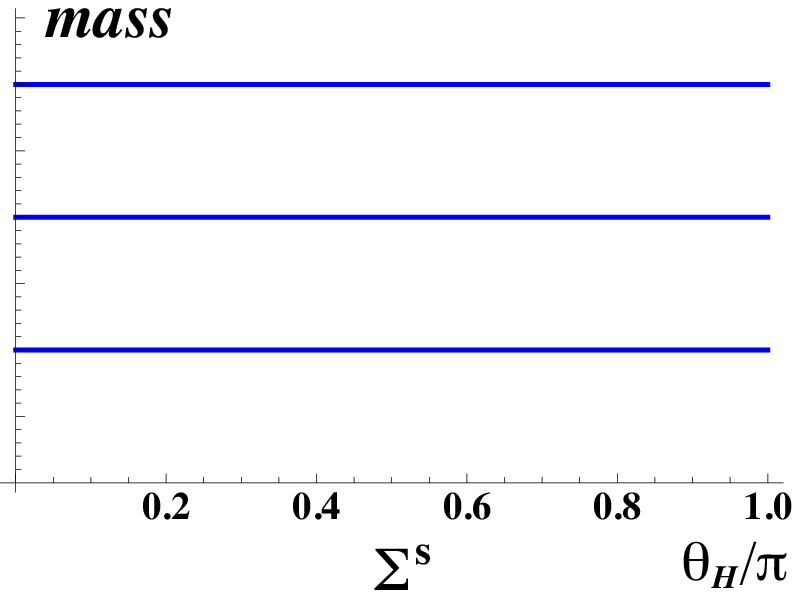}
\caption{Two types of spectra where the horizontal
axis is $\theta_H / \pi$. 
\label{fig:spec}}
\end{center}
\end{figure}

The KK expansions of the gauge fields have been worked out in ref.\ \cite{HS2}.
In the expansion on orbifolds with topology of $M^4 \times (S^1/Z_2)$, 
there appear two types of the sums.
The number of degrees of freedom on $S^1/Z_2$ is  halved compared 
with that on $S^1$.  
For those fields which acquire masses by the Hosotani mechanism
($\theta_H \not= 0$) 
two degrees of freedom combine to form one set of towers as depicted 
in Fig.\ \ref{fig:spec}
with the sum $\sum^d$.  On flat $S^1$ it corresponds to combining
cosine  and sine series for $\theta_H = 0$.  It contains a zero mode at
$\theta_H = 0$.  In the Randall-Sundrum warped space there appears a gap
in the spectrum between the two branches (corresponding the cosine and sine
series in flat space) even at $\theta_H=0$. 
The other type of a spectrum is independent of $\theta_H$, 
as depicted in Fig.\ \ref{fig:spec} with the sum $\sum^s$. 
There may or may not be a zero mode.
From the viewpoint of  the number of degrees of freedom, 
$\sum^d$ counts two KK towers, whereas $\sum^s$ counts one KK tower.

\vskip 5pt

\noindent
(i) Gauge fields

Following refs.\ \cite{HS2} and  \cite{HNU},  we expand the gauge fields in the twisted
gauge, in which $\la \tilde A_y \ra =0$, as
\beqn
\tilde{A}_\mu (x,z)  &\myeq&
    \sum_{n=0}^\infty{}^d \,
      W_\mu^{(n)}
        \left\{
          N_W (\lambda_n) 
         {T^{-_L} + T^{-_R} \over 2}
      + \cos \theta_H
        N_W (\lambda_n)
          {T^{-_L} -T^{-_R}\over 2}    \right. \cr
&& \hskip 3cm \left.     -{\sin\theta_H \over \sqrt{2}}
         D_W(\lambda_n) T^{\hat{-}} \right\}   + \textrm{h.c.} \cr
&& \hskip -1.cm
   +\sum_{n=1}^\infty {}^s \,
     W_\mu^{'(n)}
     \left\{
     -\cos \theta_H N_{W'} (\lambda_n)
       {T^{-_L} + T^{-_R} \over 2}
    + N_{W'} (\lambda_n) 
    {T^{-_L} - T^{-_R}\over 2}    \right\}    + \textrm{h.c.}   \cr
&& \hskip -1.cm
    + \sum_{n=0}^\infty {}^s \,
      A_\mu^{\gamma(n)}
        h_{\gamma}(\lambda_n)  (T^{3_L} + T^{3_R})
     + \sum_{n=1}^\infty {}^s
   A_\mu^{\hat{4}(n)} h_{A} (\lambda_n)  T^{\hat{4}}    \cr
&& \hskip -1.cm
+ \sum_{n=0}^\infty {}^d \,
    Z_\mu^{(n)}
    \bigg\{ \frac{c_\phi^2}{\sqrt{1 + s_\phi^2}}    N_Z (\lambda_n)
        {T^{3_L} + T^{3_R} \over 2}
    +\cos \theta_H
      \sqrt{1+s_\phi^2}  N_Z (\lambda_n)
        {T^{3_L} -T^{3_R}\over 2}   \cr
&& \hskip 3cm    -{\sin\theta_H \over \sqrt{2}}      \sqrt{1+s_\phi^2}
       D_Z (\lambda_n) T^{\hat{3}}        \bigg\} \cr
&& \hskip -1.cm
   + \sum_{n=1}^\infty {}^s
     Z_\mu^{'(n)}
     \left\{
       -\cos \theta_H N_{Z'}
         (\lambda_n)
     {T^{3_L} + T^{3_R} \over 2}
   + N_{Z'}
    (\lambda_n) 
    {T^{3_L}-T^{3_R} \over 2} \right\} ~, \cr
\noalign{\kern 10pt}
\tilde{B}_\mu (x,z)  &\!\!\!=\!\!\!&
  \sum_{n=0}^\infty {}^s 
  A_\mu^{\gamma (n)}
  {c_\phi \over s_\phi}
  h_{\gamma n}
 -\sum_{n=0}^\infty {}^d
  Z_\mu^{(n)} \frac{s_\phi c_\phi}{\sqrt{1+ s_\phi^2}} N_Z (\lambda_n) \cr
&& \hskip 3cm 
+ \sum_{n=1}^\infty {}^s
  Z_\mu^{'(n)} \cos \theta_H
   {s_\phi \over c_\phi}
   N_{Z'} (\lambda_n) ~ .     
\label{expansion1}
\eeqn
Here $T^{\pm} =(T^1 \pm i T^2)/\sqrt{2}$,  $c_\phi = g_A/\sqrt{g_A^2 + g_B^2}$
and $s_\phi =g_B /\sqrt{g_A^2 + g_B^2}$.
The mixing angle between $SO(5)$ and $U(1)_X$ is related to the Weinberg angle as
$\sin^2 \theta_W \equiv s_\phi^2 /(1+s_\phi^2)$.
The $n=0$ mode  stands for the zeroth mode which is massless at $\theta_H =0$.
$A_\mu^{\gamma (0)}$ remains massless at all $\theta_H$.
The $W$ and $Z$ bosons and the photon $\gamma$ correspond to
$W_\mu^{(0)}$, $Z_\mu^{(0)}$ and $A_\mu^{\gamma (0)}$, respectively.  
Unless confusion arises, we will omit the superscript $(0)$ for representing
the lowest mode.  The mode functions $N_W (\lambda) = N_W (z, \lambda)$, 
$D_W (\lambda) = D_W (z, \lambda)$ etc.\ are expressed in terms of Bessel
functions
\bea
&&\hskip -1cm 
C(z;\lambda) =
      {\pi\over 2}\lambda z z_L F_{1,0}(\lambda z,
        \lambda z_L) ~,
\quad
   C'(z;\lambda) =
      {\pi\over 2}\lambda^2 z z_L F_{0,0}(\lambda z,
        \lambda z_L) ~ ,
\cr
\noalign{\kern 5pt}
&&\hskip -1cm 
S(z;\lambda) =
      -{\pi\over 2}\lambda z  F_{1,1}(\lambda z,
        \lambda z_L) ~ ,
\quad
   S'(z;\lambda) =
      -{\pi\over 2}\lambda^2 z F_{0,1}(\lambda z,
        \lambda z_L) ~, 
        \cr
\noalign{\kern 5pt}
&&\hskip -1cm 
F_{\alpha,\beta}(u,v) =
    J_\alpha (u) Y_\beta(v)   -Y_\alpha(u) J_\beta(v) ~.
\label{BesselF1}
\eea
$N, h_\gamma  \propto C(z; \lambda)$ and $D, h_A  \propto S(z; \lambda)$ 
where proportionality  constants are given in ref.\ \cite{HNU}.  
For the photon ($\lambda_0 = 0$), $C(z; 0)$ is constant.

The mass spectrum $m_n = k \lambda_n$ of each KK tower is determined by
the corresponding eigenvalue equations:
\beqn
W_\mu^{(n)} &:& 2 S(1; \lambda_n) C' (1; \lambda_n) + \lambda_n \sin^2 \theta_H =0 ~, \cr
\noalign{\kern 5pt}
W_\mu^{'(n)} &:& C(1; \lambda_n) = 0 ~, \cr
\noalign{\kern 5pt}
Z_\mu^{(n)} &:& 
2 S(1; \lambda_n) C' (1; \lambda_n) + \lambda_n  (1+ s_\phi^2)  \sin^2 \theta_H =0 ~, \cr
\noalign{\kern 5pt}
Z_\mu^{'(n)}  &:& C(1; \lambda_n) = 0 ~, \cr
\noalign{\kern 5pt}
A_\mu^{\gamma(n)} &:& C' (1; \lambda_n) = 0 ~, \cr
\noalign{\kern 5pt}
A_\mu^{\hat{4}(n)}  &:& S(1; \lambda_n) = 0 ~.
\label{spectrum1}
\eeqn
At $\theta_H = \onehalf \pi$, the Weinberg angle $\theta_W$ is determined by global fit of
various quantities.  With $m_Z$ and $z_L$ as an input, the AdS curvature
$k$ and the $W$ boson mass at the tree level are determined 
as in Table~\ref{mH-table}. 
Counting the number of mass eigenvalue equations in (\ref{spectrum1}),
one finds  that
the 11 degrees of freedom for the original 
$SO(5)\times U(1)_X$ gauge fields
$\tilde{A}_\mu$ and $\tilde{B}_\mu$ are decomposed 
into charged components, 4 $W_\mu^{(n)}$ and 2  $W_\mu^{\prime (n)}$, 
and neutral components, 
2 $Z_\mu^{(n)}$, 1 $Z_\mu^{\prime (n)}$,
1 $A_\mu^{\gamma(n)}$ and
1 $A_\mu^{\hat{4}(n)}$.

Similarly the fifth-dimensional components $A_z$ and $B_z$ are expanded as
\beqn
&&\hskip -1.cm
\tilde{A}_z (x,z)    =
 \sum_{n=1}^\infty {}^s  \sum_{a=1}^3   
 S^{a (n)} h_{S}^{LR}(\lambda_n)  \frac{T^{a_L} + T^{a_R}}{\sqrt{2}}
+ \sum_{n=0}^\infty {}^s  H^{(n)} h_{H}^\wedge (\lambda_n)  T^{\hat{4}}    \cr
\noalign{\kern 5pt}
&&\hskip -.5cm    + \sum_{n=1}^\infty {}^d  \sum_{a=1}^3   
D^{a(n)}  \bigg\{   v_n(\theta_H, \lambda_n) h_{D}^{LR}(\lambda_n)  
\frac{T^{a_L} -T^{a_R}}{\sqrt{2}}
+w_n(\theta_H, \lambda_n) h_{D}^\wedge (\lambda_n)  T^{\hat{a}} \bigg\} ~, \cr
\noalign{\kern 10pt}
&&\hskip -1.cm
\tilde{B}_z (x,z)  = \sum_{n=1}^\infty {}^s   B^{(n)} h_{B} (\lambda_n) ~.
\label{expansion2}
\eeqn
Here $h_{S}^{LR}, h_{D}^{LR}, h_B \propto C'(z; \lambda)$ and
$h_{H}^\wedge, h_{D}^\wedge  \propto S'(z; \lambda)$.
$H(x) = H^{(0)} (x)$ is the 4D neutral Higgs boson.
The wave functions of $D^{a(n)}$ are rather involved.
The mass spectrum of each KK tower is given by
\beqn
S^{a (n)}&:& C'(1; \lambda_n) = 0 ~, \cr
\noalign{\kern 5pt}
B^{(n)} &:& C'(1; \lambda_n) = 0 ~, \cr
\noalign{\kern 5pt}
D^{a (n)}&:&  S(1; \lambda_n) C' (1; \lambda_n) + \lambda_n \sin^2 \theta_H  \cr
&& = C(1; \lambda_n) S' (1; \lambda_n) - \lambda_n \cos^2 \theta_H = 0 ~,  \cr
\noalign{\kern 5pt}
H^{(n)} &:& S(1; \lambda_n) = 0 ~, 
\label{spectrum2}
\eeqn
The 11 degrees of freedom
for the original $SO(5)\times U(1)$ gauge fields
$\tilde{A}_z$ and $\tilde{B}_z$ are decomposed into
3 $S^{(n)}$, 6 $D^{(n)}$, 1 $H^{(n)}$ and 1 $B^{(n)}$.

At $\theta_H = \onehalf \pi$ the KK expansion of $\tilde A_z$ takes a simpler
form.  The modes $\{ D^{a (n)}\}$ split into two classes;
\beqn
&&\hskip -1.cm
\tilde{A}_z (x,z)    =
 \sum_{n=1}^\infty {}^s  \sum_{a=1}^3   
 S^{a (n)} h_{S}^{LR}(\lambda_n)  \frac{T^{a_L} + T^{a_R}}{\sqrt{2}}
+ \sum_{n=0}^\infty {}^s  H^{(n)} h_{H}^\wedge (\lambda_n)  T^{\hat{4}}    \cr
\noalign{\kern 5pt}
&&\hskip -.0cm    + \sum_{n=1}^\infty {}^s  \sum_{a=1}^3   
D_-^{a(n)}   h_{D}^{LR}(\lambda_n)  \frac{T^{a_L} -T^{a_R}}{\sqrt{2}}
+ \sum_{n=1}^\infty {}^s  \sum_{a=1}^3
\hat D^{a (n)}  h_{D}^\wedge (\lambda_n)  T^{\hat{a}}~,  \cr
\noalign{\kern 10pt}
&&\hskip 1.cm
D_-^{a(n)} ~:~ C(1; \lambda_n) =0 ~, \cr
&&\hskip 1.cm
\hat D^{a (n)} ~:~ S'(1; \lambda_n) =0 ~.
\label{expansion3}
\eeqn

At this stage we recall the algebra of the generators $\{ T^\alpha \}$ of  $SO(5)$;
\beqn
&&\hskip -1cm
[T^{a_L} , T^{b_L}] = i \ep^{abc} T^{c_L} ~~,~~
[T^{a_R} , T^{b_R}] = i \ep^{abc} T^{c_R} ~~,~~
[T^{a_L} , T^{b_R}] = 0 ~~, \cr
\noalign{\kern 10pt}
&&\hskip -1cm
[T^{\hat a} , T^{\hat b}] = \frac{i}{2} \ep^{abc} (T^{c_L} + T^{c_R} ) ~~,~~ \cr
\noalign{\kern 5pt}
&&\hskip -1cm
[T^{\hat a} , T^{b_L}] = -  \frac{i}{2} \delta^{ab} T^{\hat 4} + \frac{i}{2} \ep^{abc} T^{\hat c}  ~~,~~
[T^{\hat a} , T^{b_R}] = + \frac{i}{2} \delta^{ab} T^{\hat 4} + \frac{i}{2} \ep^{abc} T^{\hat c}  ~~, \cr
\noalign{\kern 5pt}
&&\hskip -1cm
[T^{a_L} , T^{\hat 4}] = -  \frac{i}{2}  T^{\hat a} ~~,~~
[T^{a_R} , T^{\hat 4}] = +  \frac{i}{2}  T^{\hat a} ~~,~~
[T^{\hat a} , T^{\hat 4}] =   \frac{i}{2}   (T^{a_L} - T^{a_R} )  ~~, \cr
\noalign{\kern 5pt}
&&\hskip 2.cm
(a,b, c = 1 \sim 3 )~.
\label{algebra1}
\eeqn
The explicit matrix representations of $\{ T^\alpha \}$ are given in ref.\ \cite{ACP}.
The algebra remains invariant under the substitution of
$\{ T^\alpha \} = \{ T^{a_L}, T^{a_R}, T^{\hat a}, T^{\hat 4} \}$ by
$\{ T^{\prime \alpha} \}  = \{ T^{a_R}, T^{a_L}, T^{\hat a}, -T^{\hat 4} \}$.
The two sets are related to each other by an $O(5)$ transformation
$T^{\prime \alpha} = \Omega_H T^\alpha \Omega_H^{-1}$ where 
$\Omega_H = {\rm diag}\, (1,1,1,-1,1)$ in the vectorial representation.
$\Omega_H$ interchanges $SU(2)_L$ and $SU(2)_R$ and flips the direction of
$T^{\hat 4}$.

At $\theta_H = \onehalf \pi$ ($\cos \theta_H =0$) additional symmetry arises in the 
expansions.  Look at, for instance, $W_\mu^{(n)}$ and $W_\mu^{\prime (n)}$ 
terms in (\ref{expansion1}).  At $\theta_H = \onehalf \pi$ the $W_\mu^{(n)}$
terms are invariant under $\Omega_H$, whereas the $W_\mu^{\prime (n)}$
term flips the sign.
Indeed,   $\Omega_H \tilde A_M (x,z) \Omega_H^{-1}$ is the same as
$\tilde A_M (x,z)$ where the signs of the fields
\beeq
W^{\prime (n)}_\mu , Z^{\prime (n)}_\mu, A_\mu^{\hat 4 (n)}, 
H^{(n)}, D_-^{a(n)}
\quad (P_H ~ \hbox{odd})
\label{Hodd1}
\eneq
are flipped.   This defines $H$ parity ($P_H$)  for all 4D fields.  
4D fields contained in $\tilde A_M$ other than those in (\ref{Hodd1}) 
are $P_H$  even.

The action of the pure gauge fields in the bulk, $\Tr F_{MN} F^{MN}$, 
is invariant under the $\Omega_H$ transformation 
so that it is invariant under $H$ parity.  We show below that the invariance holds
for the entire action including the bulk fermions, brane fermions, and brane scalar.

\vskip 5pt

\noindent
(ii) Fermions

$H$ parity of fermions is determined in the following manner.  
Consider the fermion multiplets
containing quarks, namely, $\Psi_1$ and $\Psi_2$ in (\ref{bulkF1}) and
${\hat\chi}^q_{1R}$, ${\hat\chi}^q_{2R}$,  ${\hat\chi}^q_{3R}$,   
in (\ref{braneF1}).  They are classified in terms of electric charge 
$Q_E=\frac{5}{3}$, $\frac{2}{3}$, $-\frac{1}{3}$, $-\frac{4}{3}$.  
Recall that components of $\check \Psi$ in (\ref{bulkF1}) are related to
the components $\Psi^k$ ($k= 1 \sim 5$) in the vectorial representation by
\beeq
\check{\Psi} = 
\begin{pmatrix}
      \check{\Psi}_{11} & \check{\Psi}_{12} \cr
      \check{\Psi}_{21} & \check{\Psi}_{22} 
\end{pmatrix} 
= - \frac{1}{\sqrt{2}} 
\begin{pmatrix}
     \Psi^2 +i\Psi^1 & -\Psi^4 - i\Psi^3 \\
     \Psi^4 -i\Psi^3 &  \Psi^2 -i\Psi^1 \\
\end{pmatrix} ~.
\label{bulkF2}
\eneq
Only $\Psi^4$ and $\Psi^5$ couple with $\theta_H$.
By $\Omega_H$   the bulk fermions are transformed, in the twisted gauge,  to 
$\tilde \Psi (x, z)  \go \Omega_H \tilde \Psi (x, z)$.  In the vectorial
representation $(\tilde \Psi^1,  \tilde \Psi^2,\tilde \Psi^3,\tilde \Psi^4,\tilde \Psi^5)
 \go (\tilde \Psi^1,  \tilde \Psi^2,\tilde \Psi^3, -\tilde \Psi^4,\tilde \Psi^5)$.

The $Q_E = \frac{5}{3}$ sector consists of $T$ in $\Psi_1$ and 
$\hat T_R$ in $\hat \chi^q_{1R}$.  These fields do not couple to $\theta_H$
so that the spectrum and mode functions are independent of $\theta_H$.
The 4D fields in this sector are all  $P_H$  even.

The $Q_E = \frac{2}{3}$ sector consists of $B$, $t$, $t'$ in $\Psi_1$,
$U$  in $\Psi_2$,  $\hat B_R$ in $\hat \chi^q_{1R}$ and
$\hat U_R$ in $\hat \chi^q_{2R}$.  These fields are intertwined by $\theta_H \not= 0$.
The spectrum and wave functions of the low-lying modes have been given in
refs.\ \cite{HOOS, HK, HNU}.  The arguments can be generalized to KK modes
as well.

The boundary conditions at the TeV brane demand that
the left- and right-handed fields are expanded in the twisted  gauge  as
\beqn
\begin{pmatrix}  \tilde{U}_L \\  (\tilde{B}_L \pm \tilde{t}_L)/\sqrt{2} \\
     \tilde{t}_L'    \end{pmatrix}  (x,z)
&\myeq &  \sqrt{k}  \sum_n 
\begin{pmatrix}   a_U^{(n)}  C_L(z;\lambda_n , c_2) \\
   a_{B\pm t}^{(n)} C_L(z; \lambda_n , c_1) \\
   a_{t'}^{(n)}  S_L(z;\lambda_n, c_1) \end{pmatrix}  \psi^{(n)}_{\frac{2}{3}, L}(x) ~ , \cr
\noalign{\kern 5pt}
\begin{pmatrix}  \tilde{U}_R \\  (\tilde{B}_R \pm \tilde{t}_R)/\sqrt{2} \\
     \tilde{t}_R'    \end{pmatrix}  (x,z)
&\myeq &   \sqrt{k}  \sum_n 
\begin{pmatrix}   a_U^{(n)}  S_R(z;\lambda_n , c_2) \\
   a_{B\pm t}^{(n)} S_R(z; \lambda_n , c_1) \\
   a_{t'}^{(n)}  C_R(z;\lambda_n, c_1) \end{pmatrix}  \psi^{(n)}_{\frac{2}{3}, R}(x) ~ . 
\label{Fprofile1}
\eeqn
Here $c_a$ is the bulk kink mass for $\Psi_a$, and 
\beqn
&&\hskip -1cm
\begin{pmatrix} C_L \cr S_L \end{pmatrix}  (z;\lambda, c)
= \pm \frac{\pi}{2} \lambda\sqrt{zz_L}
   F_{c+{1\over 2},c\mp{1\over 2}}  (\lambda z, \lambda z_L) ~, \cr
\noalign{\kern 10pt}
&&\hskip -1cm  
\begin{pmatrix} C_R \cr S_R \end{pmatrix}  (z;\lambda, c)
= \mp \frac{\pi}{2} \lambda\sqrt{zz_L}
F_{c-{1\over 2},c\pm {1\over 2}} 
 (\lambda z, \lambda z_L)~.
\label{Bessel2}
\eeqn
The brane fields $\hat B_R$ and $\hat U_R$ can be expressed in terms of 
the bulk fields.  

The boundary conditions at the Planck brane lead to a matrix equation
\beeq
K ~ \begin{pmatrix}
a_U^{(n)} \cr  \myfrac{1}{2}~ a_{B + t}^{(n)}\cr   
\myfrac{1}{\sqrt{2}} ~ a_{t'}^{(n)} \cr
\myfrac{1}{2} ~ a_{B - t}^{(n)}  
\end{pmatrix} = 0 ~, 
\label{Feq1}
\eneq
where 
\beqn
&&\hskip -1.cm
K= \cr
\noalign{\kern 15pt}
&&\hskip -1.cm
\begin{pmatrix}
0 & 0 & - 2 c_H S_L^{(1)} & 2s_H C_L^{(1)}  \cr
\lambda_n S_R^{(2)}- \myfrac{\mu_2^2}{2k} C_L^{(2)}
& - \myfrac{\mu_2 \tilde\mu}{2k} C_L^{(1)} 
&s_H \myfrac{\mu_2 \tilde\mu}{2k} S_L^{(1)} 
&\myfrac{\mu_2 \tilde\mu}{2k} c_H C_L^{(1)} \cr
0 &  \lambda_n S_R^{(1)} - \myfrac{\mu_1^2}{2k} C_L^{(1)} 
&s_H \big( \lambda_n C_R^{(1)} - \myfrac{\mu_1^2}{2k} S_L^{(1)} \big) 
&c_H \big( \lambda_n S_R^{(1)} - \myfrac{\mu_1^2}{2k} C_L^{(1)} \big) \cr
- \myfrac{\mu_2 \tilde\mu}{2k} C_L^{(2)}
& \lambda_n S_R^{(1)} - \myfrac{\tilde\mu^2}{2k} C_L^{(1)} 
&- s_H \big( \lambda_n C_R^{(1)} - \myfrac{\tilde\mu^2}{2k} S_L^{(1)} \big) 
&- c_H \big( \lambda_n S_R^{(1)} - \myfrac{\tilde\mu^2}{2k} C_L^{(1)} \big)
\end{pmatrix} . \cr
\noalign{\kern 10pt}
&&
\label{Feq2}
\eeqn
Here $s_H = \sin \theta_H, c_H = \cos \theta_H$,  
$C_{L,R}^{(j)} = C_{L,R}(1; \lambda_n, c_j)$ and
$S_{L,R}^{(j)} = S_{L,R} (1; \lambda_n, c_j)$.

Nontrivial solutions exist with $\det K =0$, which determines the spectrum.
At $\theta_H = \onehalf \pi$, special structure appears.  Eq.\ (\ref{Feq1}) leads to
\beeq
C_{L}(1; \lambda_n, c_1)  a_{B - t}^{(n)}  = 0 ~.
\label{Feq3}
\eneq
Solutions with $C_{L}(1; \lambda_n, c_1) =0$ have  $a_{B - t}^{(n)}  \not=  0$
and $ a_{U}^{(n)} = a_{B + t}^{(n)}= a_{t'}^{(n)}=0$.
The corresponding 4D fermion tower is denoted as $\psi_{B-t}^{(n)} (x)$.
As the component $\tilde B- \tilde t $ is $\tilde \Psi_1^4$, it flips the sign under the $\Omega_H$
transformation.   The KK tower $\psi_{B-t}^{(n)} (x)$ is $P_H$  odd.  
We note that the mode function of the left-handed $\psi_{B-t,L}^{(n)} (x)$ vanishes at $z=1$, 
while that of the right-handed $\psi_{B-t,R}^{(n)} (x)$ is non-vanishing 
as seen from (\ref{Fprofile1}).
For  $C_{L}(1; \lambda_n, c_1) \not= 0$,   $a_{B - t}^{(n)} =  0$.  The spectrum and 
mode functions are determined by the 3-by-3 matrix equation reduced from (\ref{Feq1}).
They give three KK towers of 4D fermions,  including the tower of the top quark.
As $\tilde B + \tilde t  \sim \tilde \Psi_1^3$, $\tilde t' \sim \tilde \Psi_1^5$ and
$\tilde U \sim \tilde \Psi_2^2 + i \tilde \Psi_2^1$,  
these  three KK towers are all $P_H$ even.
The brane fermions $\hat B_R$ and $\hat U_R$ are related to the bulk fermions by
\beqn
&&\hskip -1.cm
\myfrac{\mu_1}{2} \hat B_R = B_R \big|_{z=1} = 
\myfrac{1}{2} (\tilde B _R + \tilde t_R) + \frac{1}{\sqrt{2}} \, \tilde t_R' ~ \bigg|_{z=1} ~, \cr
\noalign{\kern 5pt}
&&\hskip -1.cm
\myfrac{\mu_2}{2} \hat U_R = U_R \big|_{z=1} = \tilde U_R \big|_{z=1} ~.
\label{braneF2}
\eeqn
They contain only $P_H$ even fields.

Parallel arguments apply to the $Q_E = - \frac{1}{3}$ sector, which consists of 
$b$ in $\Psi_1$, $D, X, b'$ in $\Psi_2$, $\hat D_R$ in $\hat \chi^q_{2R}$ and
$\hat X_R$ in $\hat \chi^q_{3R}$.  The bulk fields are expanded as
\beqn
\begin{pmatrix}  \tilde{b}_L \\  (\tilde{D}_L \pm \tilde{X}_L)/\sqrt{2} \\
     \tilde{b}_L'    \end{pmatrix}  (x,z)
&\myeq &  \sqrt{k}  \sum_n 
\begin{pmatrix}   a_b^{(n)}  C_L(z;\lambda_n , c_1) \\
   a_{D\pm X}^{(n)} C_L(z; \lambda_n , c_2) \\
   a_{b'}^{(n)}  S_L(z;\lambda_n, c_2) \end{pmatrix}  \psi^{(n)}_{-\frac{1}{3}, L}(x) ~ , \cr
\noalign{\kern 5pt}
\begin{pmatrix}  \tilde{b}_R \\  (\tilde{D}_R \pm \tilde{X}_R)/\sqrt{2} \\
     \tilde{b}_R'    \end{pmatrix}  (x,z)
&\myeq &   \sqrt{k}  \sum_n 
\begin{pmatrix}   a_b^{(n)}  S_R(z;\lambda_n , c_1) \\
   a_{D\pm X}^{(n)} S_R(z; \lambda_n , c_2) \\
   a_{b'}^{(n)}  C_R(z;\lambda_n, c_2) \end{pmatrix}  \psi^{(n)}_{-\frac{1}{3}, R}(x) ~ . 
\label{Fprofile2}
\eeqn
The equations and relations in the  $Q_E = - \frac{1}{3}$ sector are obtained
from those in the  $Q_E = \frac{2}{3}$ sector by replacing
$(U, B, t, t')$ and $(c_1, c_2, \mu_1, \mu_2, \tilde \mu)$
by $(b, D, X, b')$ and  $(c_2, c_1, \mu_3, \tilde\mu, \mu_2)$, respectively.

At $\theta_H = \onehalf \pi$
\beeq
C_{L}(1; \lambda_n, c_2)  a_{D - X}^{(n)}  = 0 ~.
\label{Feq4}
\eneq
Solutions with $C_{L}(1; \lambda_n, c_2) =0$ have  $a_{D - X}^{(n)}  \not=  0$
and $ a_{b}^{(n)} = a_{D + X}^{(n)}= a_{b'}^{(n)}=0$.
The corresponding 4D fermion tower  denoted as $\psi_{D-X}^{(n)} (x)$
is $P_H$ odd.  The other three KK towers are $P_H$ even.
The mode function of the left-handed $\psi_{D-X,L}^{(n)} (x)$ vanishes at $z=1$, 
while that of the right-handed $\psi_{D-X,R}^{(n)} (x)$ is non-vanishing 

Finally the $Q_E = - \frac{4}{3}$ sector consisting of $Y$ in $\Psi_2$ and 
$\hat Y_R$ in $\hat \chi^q_{3R}$ does not couple to $\theta_H$.
The associated KK tower is $P_H$ even.

To summarize the 4D  fermion fields with $P_H$ odd in the third generation
are
\beeq
\psi_{B-t}^{(n)} (x),~  \psi_{D-X}^{(n)} (x), ~ 
\psi_{\tau- L_{1X}}^{(n)} (x),~  \psi_{L_{2Y}-L_{3X}}^{(n)} (x) \quad (P_H ~ \hbox{odd}).
\label{Hodd2}
\eneq
The brane fermions contain only $P_H$ even fields.

\bigskip

\noindent
(iii) $P_H$ invariance 

The bulk action (\ref{action1}) is invariant under the  $\Omega_H$ transformation, 
in which $\tilde A_M \go \Omega_H \tilde A_M \Omega_H^{-1}$ and 
$\tilde \Psi_a \go \Omega_H  \tilde \Psi_a$ in the twisted gauge.
At $\theta_H = \onehalf \pi$,  
the $P_H$ odd fields flip the sign under the transformation, 
while the $P_H$ even fields remain unaltered.  
In other words the bulk action (\ref{action1}) is invariant under the $H$ parity, $P_H$.  
The gauge fields  given in (\ref{Hodd1}),   the fermions given in (\ref{Hodd2}) 
and the corresponding ones in the first and second generations are $P_H$ odd.
All other 4D fields are $P_H$ even.

As for the brane action (\ref{action2}), we recognize first that $A_\mu^{{\hat 4} (n)}$, 
$H^{(n)}$ and $D_-^{a(n)}$ do not couple to the brane fields from the gauge invariance.  
The covariant derivative to $\Phi$ in (\ref{action2}), at first sight, seems to contain
$W_\mu^{\prime (n)}$ and $Z_\mu^{\prime (n)}$.  However, the mode functions of
$W_\mu^{\prime (n)}$ and $Z_\mu^{\prime (n)}$ are given by
$C(z; \lambda_n)$ up to proportionality constants and the spectrum is determined by
$C(1; \lambda_n) = 0$ as shown in (\ref{spectrum1}).  As a consequence
the mode functions vanish at the Planck brane and 
$W_\mu^{\prime (n)}$ and $Z_\mu^{\prime (n)}$ do not couple to the brane fields.

Similarly the mode function of the left-handed component of
$\psi_{B-t}^{(n)} (x)$ ($\psi_{D-X}^{(n)} (x)$) is given by
$C_L(z; \lambda_n, c_1)$ ($C_L(z; \lambda_n, c_2)$).
As $C_L(1; \lambda_n, c_1)=0$ ($C_L(1; \lambda_n, c_2)=0$),
the mode function vanishes at the Planck brane.
Consequently the left-handed components of 
$\psi_{B-t}^{(n)} (x)$  and $\psi_{D-X}^{(n)} (x)$ do not
appear  in the $\kappa \hat \chi^\dagger \check\Psi_L \Phi$ couplings in  (\ref{action2}).
The right-handed components of $\psi_{B-t}^{(n)} (x)$  and $\psi_{D-X}^{(n)} (x)$ do not
appear as the  brane fermions are all right-handed.

We have shown that the $P_H$ odd fields do not  couple to the brane fields.
We conclude that the total action is invariant under $P_H$.

\section{Summary}

We have shown that in the $SO(5) \times U(1)$ gauge-Higgs unification
model the energy is minimized at $\theta_H=\onehalf \pi$
and the $H$ parity ($P_H$) invariance emerges.  The parity is assigned to all
4D fields including KK excited states.  Among low energy fields only the 
4D Higgs boson is $P_H$ odd, while the quarks, leptons, $W$, $Z$, photon and
gluons are $P_H$ even.   The lowest mass among $P_H$ odd fields other than 
the Higgs boson  is of order $m_\KK$ where $m_\KK =840 \sim 1470\,$GeV
for $z_L= 10^5 \sim 10^{15}$. 

The action is invariant under $P_H$.  It is important that all bulk fermions
belong to the vector representation of $SO(5)$.  By examining the wave functions
in the fifth dimension of 4D modes and utilizing the $O(5)$ invariance in the bulk 
we have shown the invariance of the bulk action.  The fermion and scalar 
fields localized on the Planck brane couple only to $P_H$ even fields.

It follows that the Higgs boson becomes absolutely stable.  
Its consequences in cosmology and astrophysics  and in collider experiments
need to be explored further.  We will come back to them separately. 

\vskip .5cm

\leftline{\bf Acknowledgement}

This work was supported in part 
by  Scientific Grants from the Ministry of Education and Science, 
Grant No.\ 20244028 (Y.H., N.U.),  
 Grant No.\ 21244036 (Y.H.), and Grant No.\ 20244037 (M.T.).

\vskip 1cm


\renewenvironment{thebibliography}[1]
         {\begin{list}{[$\,$\arabic{enumi}$\,$]}  
         {\usecounter{enumi}\setlength{\parsep}{0pt}
          \setlength{\itemsep}{0pt}  \renewcommand{\baselinestretch}{1.2}
          \settowidth
         {\labelwidth}{#1 ~ ~}\sloppy}}{\end{list}}

\def\jnl#1#2#3#4{{#1}{\bf #2},  #3 (#4)}

\def\Zphys{{\em Z.\ Phys.} }
\def\jssc{{\em J.\ Solid State Chem.\ }}
\def\jpsJ{{\em J.\ Phys.\ Soc.\ Japan }}
\def\ptps{{\em Prog.\ Theoret.\ Phys.\ Suppl.\ }}
\def\PTP{{\em Prog.\ Theoret.\ Phys.\  }}

\def\JMP{{\em J. Math.\ Phys.} }
\def\NPB{{\em Nucl.\ Phys.} B}
\def\NP{{\em Nucl.\ Phys.} }
\def\PLB{{\it Phys.\ Lett.} B}
\def\PL{{\em Phys.\ Lett.} }
\def\PRL{\em Phys.\ Rev.\ Lett. }
\def\PRB{{\em Phys.\ Rev.} B}
\def\PRD{{\em Phys.\ Rev.} D}
\def\PRe{{\em Phys.\ Rep.} }
\def\AP{{\em Ann.\ Phys.\ (N.Y.)} }
\def\RMP{{\em Rev.\ Mod.\ Phys.} }
\def\ZPC{{\em Z.\ Phys.} C}
\def\SCI{\em Science}
\def\CMP{\em Comm.\ Math.\ Phys. }
\def\MPLA{{\em Mod.\ Phys.\ Lett.} A}
\def\IJMPA{{\em Int.\ J.\ Mod.\ Phys.} A}
\def\IJMPB{{\em Int.\ J.\ Mod.\ Phys.} B}
\def\EPJC{{\em Eur.\ Phys.\ J.} C}
\def\PR{{\em Phys.\ Rev.} }
\def\JHEP{{\em JHEP} }
\def\JCAP{{\em JCAP} }
\def\cmp{{\em Com.\ Math.\ Phys.}}
\def\JPA{{\em J.\  Phys.} A}
\def\JPG{{\em J.\  Phys.} G}
\def\NJP{{\em New.\ J.\  Phys.} }
\def\CQG{\em Class.\ Quant.\ Grav. }
\def\ATMP{{\em Adv.\ Theoret.\ Math.\ Phys.} }
\def\ibid{{\em ibid.} }


\end{document}